\documentclass[12pt]{article}

\usepackage{amsmath}
\usepackage{amsfonts}
\usepackage{amssymb}
\usepackage{amsthm}
\usepackage{enumitem}
\usepackage{fancyhdr}
\usepackage{subfig}
\usepackage{float}
\usepackage{bbm}
\usepackage{ulem}
\usepackage{setspace}
\usepackage{xcolor}
\usepackage[export]{adjustbox}
\usepackage{natbib}
\usepackage[linesnumbered,ruled]{algorithm2e}

\usepackage{graphicx}
\usepackage[colorlinks=true,linkcolor=blue,citecolor=blue]{hyperref}%

\makeatletter
\g@addto@macro{\UrlBreaks}{\UrlOrds}
\makeatother

\theoremstyle{definition}

\newtheorem{definition}{Definition}[section]

\newcommand{\blind}{1}
\begin{document}

\def\spacingset#1{\renewcommand{\baselinestretch}%
{#1}\small\normalsize} \spacingset{1}

\if1\blind
{
  \baselineskip=28pt \vskip 5mm
    \begin{center} {\LARGE{\bf Scalable Multiple Changepoint Detection for Functional Data Sequences}}
    \end{center}

    \baselineskip=14pt \vskip 10mm

    \begin{center}\large
    Trevor Harris \footnote{\baselineskip=12pt Department of Statistics, University of Illinois at Urbana-Champaign} \footnote{\baselineskip=12pt Corresponding Author - Trevor Harris: trevorh2@illinois.edu, Bo Li: libo@illinois.edu, J. Derek Tucker: jdtuck@sandia.gov},
        Bo Li$^1$,
        J. Derek Tucker$^{1,}$\footnote{\baselineskip=12pt Sandia National Laboratories, Albuquerque, NM}
    \end{center}
    \baselineskip=19pt \vskip 15mm \centerline{\today} \vskip 6mm
} \fi

\if0\blind
{
  \bigskip
  \bigskip
  \bigskip
  \begin{center}
    {\LARGE\bf Scalable Multiple Changepoint Detection for Functional Data Sequences}
\end{center}
  \medskip
} \fi

\begin{center}
{\large{\bf Abstract}}
\end{center}

We propose the Multiple Changepoint Isolation (MCI) method for detecting multiple changes in the mean and covariance of a functional process. We first introduce a pair of projections to represent the variability ``between'' and ``within'' the functional observations. We then present an augmented fused lasso procedure to split the projections into multiple regions robustly. These regions act to isolate each changepoint away from the others so that the powerful univariate CUSUM statistic can be applied region-wise to identify the changepoints. Simulations show that our method accurately detects the number and locations of changepoints under many different scenarios. These include light and heavy tailed data, data with symmetric and skewed distributions, sparsely and densely sampled changepoints, and mean and covariance changes.  We show that our method outperforms a recent multiple functional changepoint detector and several univariate changepoint detectors applied to our proposed projections. We also show that MCI is more robust than existing approaches and scales linearly with sample size. Finally, we demonstrate our method on a large time series of water vapor mixing ratio profiles from atmospheric emitted radiance interferometer measurements.

\baselineskip=14pt

\par\vfill\noindent

\par\medskip\noindent
{\bf Short title}: Functional Changepoint Detection

\clearpage\pagebreak \pagenumbering{arabic}
\newpage \baselineskip=24pt


\section{Introduction} \label{sec:intro}

The statistical analysis of functional time series has become increasingly important to many scientific fields including Climatology \citep{shang2011nonparametric}, Finance \citep{kokoszka2012functional}, Geophysics \citep{hormann2012functional}, Demography \citep{hyndman2008stochastic}, Manufacturing \citep{woodall2007current}, 
and environmental modeling \citep{finazzi2019statistical,fortuna2020functional, qu2021robust}. 
A functional time series is a sequence of infinite dimensional objects, such as curves and surfaces, observed over time. Functional time series are analogous to univariate or multivariate time series, except that we observe a continuous function at each point in time \citep{bosq2012linear}.
Just as in univariate and multivariate time series, a functional time series can experience abrupt changes in its generating process.
These abrupt changes, or changepoints, can complicate statistical analysis by invalidating stationarity assumptions. However, they can also be interesting in their own right by revealing unexpected heterogeneous patterns. 

Recently, changepoint analysis has become particularly important in climatological and environmental process modeling  \citep{lee2020trend, jaruvskova2020changepoint, beaulieu2020considering}. 
Although these works focus on important univariate changepoint detection applications, environmental profiles, such as water vapor columns \citep{sakai2019automated}, are increasingly the object of study. Thus changepoint detection in functional time series are a natural and timely extension to the finite dimensional methods.

Within the Functional Data Analysis (FDA) literature, changepoint detection has largely focused on the At Most One Change (AMOC) problem. In \cite{berkes2009detecting} a Cumulative Sum (CUSUM) test was proposed for independent functional data, which was further studied in \cite{aue2009estimation}, where its asymptotic properties were developed. This test was then extended to weakly dependent functional data by \cite{hormann2010weakly} and epidemic changes by \cite{aston2012detecting}. \cite{zhang2011testing} introduced a test for changes in the mean of weakly dependent functional data using self-normalization to alleviate the use of asymptotic control. Later, \cite{sharipov2016sequential} similarly developed a sequential block bootstrap procedure for these methods. Recently, \cite{gromenko2017detection} considered changes in spatially correlated functional data, and \cite{aue2018detecting} proposed a fully functional method for finding a mean change without losing information due to dimension reduction.

Detecting multiple changepoints in a functional time series has received relatively scant attention compared to the AMOC problem. Recently, \cite{li2018bayesian} proposed a Bayesian method for identifying multiple changepoints in the mean by transforming the functional data into wavelets and identifying changes in the wavelet coefficient processes. In addition, \cite{chiou2019identifying} introduced a dynamic segmentation method for multiple changepoints in the mean based on dynamic programming and backward elimination to find an optimal set of changepoints. Multiple changepoints can also be identified by augmenting AMOC methods with a recursive binary segmentation algorithm to partition the functional time series \citep{berkes2009detecting, aue2018detecting}. The consistency of binary segmentation approaches was established by \citet{rice2019consistency}. Despite these advances, there are several outstanding issues with these approaches that we hope to address. Namely, sub-optimal computational scalability, insufficient power to detect covariance and shape based alternatives, and a lack of robustness.

The computational complexity of functional multiple changepoints detection methods has been an obstacle for their wide application to large functional time series. Bayesian methods that rely on Markov Chain Monte Carlo sampling are intrinsically burdensome because they typically require an enormous number of samples to reach convergence.
Ordinary dynamic programming and binary segmentation algorithms scale quadratically and log-linearly, respectively, with the data's sample size.
As larger and larger functional time series data sets are curated, methods that scale linearly with sample size are called for to meet the computational demand.
In the univariate and multivariate changepoint detection literature, computationally efficient methods for multiple changepoint detection have already emerged. 
These include the cumulative segmentation method \citep{muggeo2011efficient},
the Pruned Exact Linear Time (PELT) algorithm \citep{killick2012optimal}, the Functional Pruning Optimal Partitioning (FPOP) algorithm \citep{maidstone2017optimal}, and the robust Functional Pruning Optimal Partitioning (r-FPOP) \citep{fearnhead2019changepoint}.

Another limitation of the existing functional changepoint detection methods is that most methods can only detect changes in the functional process's mean. While mean changes are the most conspicuous, covariance changes are equally important and can also occur. The need for detecting covariance changes has already been noticed and tackled in the univariate time series literature, where methods targeting 
variance changes have been developed \citep{adelfio2012change, chapman2020nonparametric}. Recently, covariance changepoint detection has been considered in functional data \citep{aue2020structural},
however, methods that target both mean and covariance changes in a functional process are still not available. 
Lastly, many previous methods were developed under the assumption that the data follows a Gaussian process. Their performance on non-Gaussian, skewed, or heavy tailed data, which may be encountered in practice, is not well studied and could potentially be suboptimal.

%

We propose the Multiple Changepoint Isolation (MCI) method to robustly detect multiple changepoints in a functional time series' distribution. Our method uses a pair of projections to minimally represent the variability ``between'' and ``within'' individual functions. Although there is no clear direct connection between the projections and specific parameters, we show that the projections largely preserve the mean and covariance information (Section \ref{sec:simulations})
We then detect changepoints in each projection using an augmented fused lasso procedure based on the CUSUM statistic \citep{page1954continuous}. After that, changepoints are aggregated across all projections into a final list.
This approach combines robust segmentation through the fused lasso \citep{tibshirani2005sparsity} with optimal detection through the CUSUM to achieve very low error rates compared with existing methods (see Sections \ref{sec:simulations} and supplement Section B). 

Our method also remedies three outstanding issues in the functional changepoint detection literature. First, our method detects a broader range of changepoint types, beyond mean changes, than existing methods. Second, our method is computationally efficient, having only linear time computational complexity.
Finally, our method is robust to asymmetry and heavier than Gaussian tails. 

\section{Multiple Changepoint Isolation} \label{sec:methods}

Let $\{f_t, t \in 1,...,n \}$ be a sequence of continuous functions in $L^2([0, 1])$, hereafter $L^2$, the Hilbert space of square integrable functions with domain $[0, 1]$ satisfying $\int_0^1 |f_t(s)|^2ds < \infty$. The domain $[0, 1]$ is assumed without loss of generality, and we further assume that each function $f_t \in \{f_t, t \in 1,...,n \}$ is observed on the same finite grid of points $0 < s_1 < ... < s_m < 1$. 
Suppose that the observations $\{f_t, t \in 1,...,n \}$ are distributed according to
\begin{equation} \label{eqn:cp_model}
    f_t \sim F(\mu_t, \Sigma_t),
\end{equation}
where $F(\mu_t, \Sigma_t)$ refers to a functional process with mean function $\mu_t$ and covariance function $\Sigma_t$. We assume that $F(\cdot)$ is piecewise weakly stationary, that is $\mu_{t+1} = \mu_{t}$ and $\Sigma_{t+1} = \Sigma_{t}$ for all $t \in 1,...,n$ except when $t$ is a changepoint. 
We assume there are $M_1$ time points where $\mu_{t+1} \neq \mu_{t}$ and $M_2$ time points where $\Sigma_{t+1} \neq \Sigma_{t}$. Both $M_1$ and $M_2$ are unknown integers between 1 and $n$. 
Let $\tau_j^{\mu}$ with $j = 1,...,M_1$ and $\tau_k^{\Sigma}$ with $k = 1,...,M_2$ denote the times where $\mu_t$ and $\Sigma_t$ change, i.e., where $\mu_{\tau_j^\mu+1} \neq \mu_{\tau_j^\mu}$ and $\Sigma_{\tau_k^\Sigma+1} \neq \Sigma_{\tau_k^\Sigma}$. For simplicity and ease of notation, we union the two sets of changepoints into a combined sequence $\tau_1,...,\tau_M$, where $\max(M_1, M_2) \le M < n$.
The goal of our method is to estimate \textit{all} changepoint locations $\tau_1,...,\tau_M$ from the sequence $\{f_t, t \in 1,...,n \}$.

\subsection{Main algorithm} \label{sec:main_alg}
We divide our detection algorithm into four distinct steps. The first two are proposed to give initial estimates of changepoints so that the third step can finely tune the initial estimates using the standard CUSUM statistic with a false discovery rate control. The final step aggregates all changepoints detected across all projections together. We specify the four steps of our detection algorithm below: 

\begin{enumerate}
    \item {\bf Univariate projection.} Project the functional data, $\{f_t, t \in 1,...,n \}$ onto $\mathbbm{R}$ with two projections, 
  one representing variability ``between'' different curves, denoted by $P_T$, and the other representing variability ``within'' individual curves, denoted by $P_{E1}$. The two projections yield two projected time series, $Y^{(1)} = \{Y^{(1)}_t, t \in 1,\dots, n\}$ and $Y^{(2)} = \{Y^{(2)}_t, t \in 1,\dots, n\}$.
    \item {\bf Changeset regionalization.} Represent each projected process as $Y^{(i)} = \theta^{(i)} + \epsilon^{(i)}$ for $i=1,2$, where $\theta^{(i)}$ is a piecewise constant vector with entries $\theta_t^{(i)}$ and $\epsilon^{(i)}$ is \textit{i.i.d} noise. Estimate $\hat{\theta^{(i)}} = \arg\min_{\theta}||Y^{(i)} - \theta^{(i)}||_2^2 + \lambda\sum_{t=1}^{n-1} |\theta^{(i)}_{t+1} - \theta^{(i)}_t|$, i.e. with the fused lasso estimator. Group nearby jumps in  $\hat{\theta^{(i)}}$ within $c$ time steps into individual sets, called ``changesets''. The changesets define subregions of $Y^{(i)}$, with at least $2c + 1$ data points each, that are assumed to contain AMOC. 
    \item {\bf CUSUM.} Apply CUSUM regionwise to detect the potential changepoint in each subregion of the two projections. Apply a false discovery rate correction to control the number of false detections at a specified level $\alpha$. Retain the changepoints whose p-values after correction are below $\alpha$. Let $\hat{\tau}^{(1)}_1,...,\hat{\tau}^{(1)}_{m_1}$ and $\hat{\tau}^{(2)}_1,...,\hat{\tau}^{(2)}_{m_2}$ denote the estimated changepoints from the two projections, respectively. 
    \item {\bf Finalization.} Concatenate, sort, and deduplicate the retained changepoints from both projections into a single list, denoted by $\hat{\tau}_1,...,\hat{\tau}_{m}$, which are our final estimates of the true changepoints.
    
\end{enumerate}

Three hyper-parameters are involved in the algorithm: the fused lasso penalty term $\lambda$, the linkage parameter $c$, and the false positive rate $\alpha$. We use the data to optimize the $\lambda$ and $c$ parameters through grid search. Details are given in Section \ref{sec:bic_optimzation}. The false positive level $\alpha$ is considered fixed, as is common in hypothesis testing, although this parameter could be optimized as well. A small simulation study in Section B.2 in the supplement shows that MCI is relatively insensitive to the prespecified $\alpha$ level, as long as $\alpha$ is below $0.01$.
We elaborate the first two steps of our algorithm in the following two subsections. 

\subsection{Univariate projections} \label{sec:projections}

Step one projects the functional data sequence $\{f_t, t \in 1,...,n \}$ onto $\mathbbm{R}$ through two projections $P_T$ and $P_{E1}$,
which are defined as
\begin{align}
    P_T(f_t) &= \int_0^1 |\nabla f_t(s)| ds, \\
    P_{E1}(f_t) &= \int_0^1 f_t(s) \phi_1(s) ds,
\end{align}
where $\nabla$ is the differential operator, $|\cdot|$ is the $L^2$ norm, and $\phi_1$ is the eigenfunction of the covariance operator of $f_t$ corresponding to the largest singular value.

The first projection $P_T$ is a total variation norm (TVN) type projection of $f_t$. The TVN based projection measures the variability ``within'' the function $f_t$ by measuring the total squared length of $f_t$, and it is invariant to rigid and elastic transformations. The projection $P_T$ is sensitive to the high-frequency features in  $\{f_t, t \in 1,...,n \}$, because features at higher frequencies contribute more to the length than lower frequencies.
Therefore, $P_T$ can be more helpful in detecting changes in the covariance operators $\Sigma_t$ (Equation \ref{eqn:cp_model}), such as changes in variance or smoothness of the functions (Section \ref{sec:assess_results}). 

The second projection $P_{E1}$ is the first functional principal component, which measures the dominant mode of variation in the data and can be estimated through standard Functional Principal Component Analysis \citep{ramsay2004functional} or elastic Functional Principal Component Analysis if the data contains phase variability \citep{tucker2013generative,srivastava2016functional}.
Whereas $P_T$ is sensitive to high frequency features, $P_{E1}$ is sensitive to low frequency features such as the mean, slope, orientation, and sign of functional curves. Combining $P_T$ and $P_{E1}$ will retain the information in functional sequences at both high and low frequencies. 

We could include further FPC coefficients, say $P_{E2}(f_t) = \int_0^1 f_t(s) \phi_2(s) ds$ and $P_{E3} (f_t) = \int_0^1 f_t(s) \phi_3(s) ds$, but we find this is unnecessary since $P_T$ is more sensitive to high frequency changes than any individual FPC (Section B.2 of the supplement). Using only the top FPC also avoids having to select the right number of FPCs through a potentially expensive cross-validation procedure \citep{hyun2016exact}.

\begin{figure}
  \centering
  \includegraphics[width=0.85\textwidth]{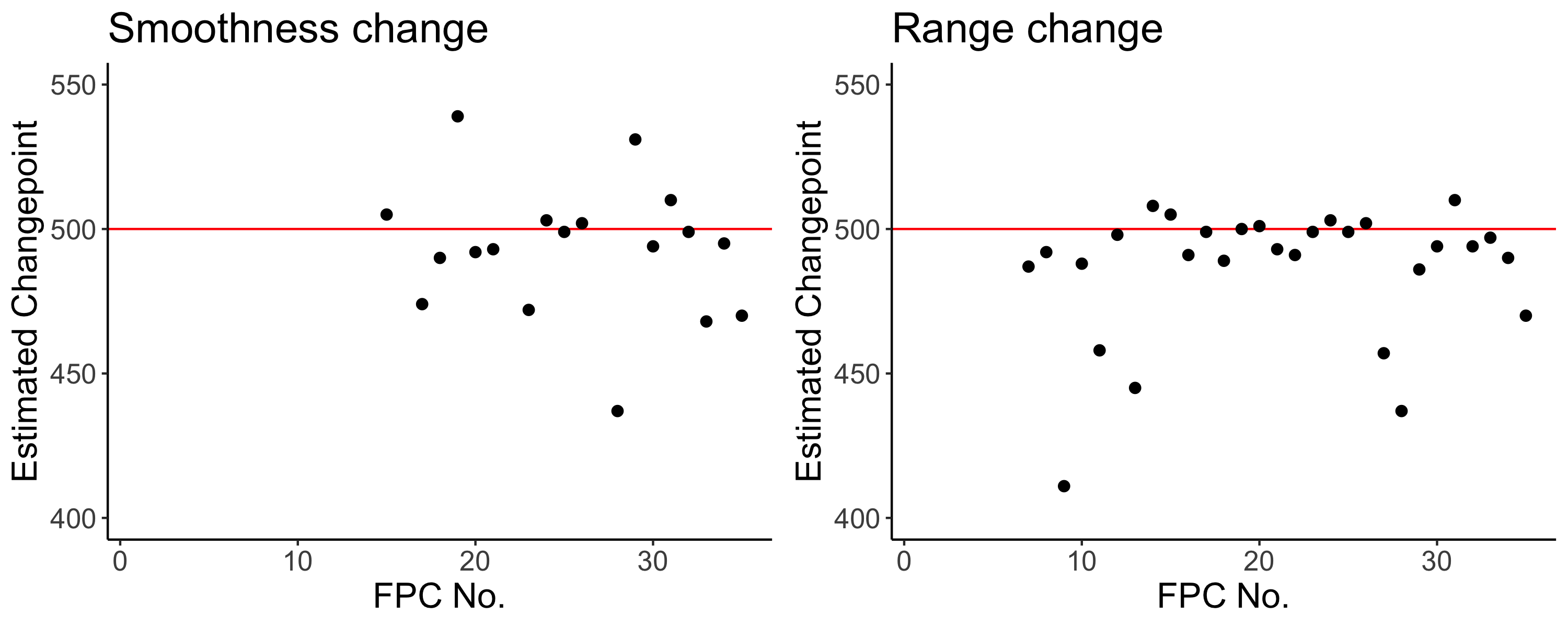}
  \caption{Estimated changepoints, using CUSUM applied to each FPC, under two types of changes (smoothness and range -- see Equation \ref{eqn:matern}) for a functional time series with $n = 1000$ and one changepoint at $t = 500$. CUSUM estimates (black dots) are generally centered around the true changepoint location (red line) but show high variability. No black dot indicates that no changepoint was detected in that FPC. This illustrates a major challenge in detecting changepoints, particularly covariance-based changepoints, purely through FPC analysis since the estimate can vary wildly depending on which FPC is used.}
  \label{fig:fpc_bad}
\end{figure}

Univariate projections are a classic and powerful technique in functional data for extracting features and compressing a functional process to a finite dimensional space. Our system of only two projections is based on the principle that the number of projections should be minimized and that each projection should provide a high amount of independent information.
Overly projecting the data can lead to two adverse effects in change point detection: increased estimation error and reduced test power. Estimation error increases because changes will often partially manifest in multiple projections simultaneously, meaning changepoints may be repeatedly estimated on noisy representations. This can lead to many spurious changepoint estimates near the true ones (Figure \ref{fig:fpc_bad}). Test power decreases because more projections mean more tests are being performed, so the necessary false discovery rate or family wise error rate correction will have to control for a higher number of tests. This directly leads to a power loss for each test, which can result in missing some changepoints entirely.

\subsection{Changeset regionalization} \label{sec:changeset_regionalization}

Let $Y^{(i)} = \{Y^{(i)}_t, t \in 1,\dots, n\}$, $i=1,2$, denote the two projections of the functional sequence $\{f_t, t \in 1,...,n \}$ by $P_T$ and $P_{E1}$. Each projection conveniently converts changes in the spectrum of $f_t$ into simple mean changes in $Y_t^{(i)}$. We model the mean changes, or jumps, with a signal-plus-noise model, $Y^{(i)} = \theta^{(i)} + \epsilon^{(i)}$, where $\theta^{(i)}$ is a piecewise constant $n\times1$ vector with entries $\theta^{(i)}_t$, and the $i.i.d.$ noise term $\epsilon^{(i)}$ has a finite second moment. Because $\theta^{(i)}$ is piecewise constant, the changepoints in $Y^{(i)}$ are reflected in the jumps, or non-zero differences, of $\theta^{(i)}$, i.e. where $\theta^{(i)}_{t+1} \neq \theta^{(i)}_t$. Without any risk of confusion, we suppress the superscript $(i)$ to ease the notation in what follows. 

Ideally, we could estimate $\hat{\theta}$ with an $L^0$ norm penalized empirical risk estimator, since $L^0$ minimization asymptotically recovers the jump locations \citep{boysen2009consistencies}.
However, $L^0$ minimization is often computationally infeasible. We, therefore, instead start with a convex approximation, $L^1$ minimization, and try to modify the approximation into an $L^0$-like solution.
The $L^1$ minimization, in this case known as the fused lasso \citep{tibshirani2005sparsity}, has received a flurry of recent interest \citep{rojas2014change, chan2014group, hyun2016exact, lin2016approximate} for changepoint estimation. 
The fused lasso is an optimally adaptive solution to piecewise signal recovery \citep{mammen1997locally, donoho1998minimax, guntuboyina2017adaptive} that can be solved for with non-iterative convex optimization \citep{condat2013direct}. Thus it can be used to accurately and quickly approximate $\theta$.
The fused lasso estimates $\theta$ via
\begin{align} \label{eqn:l0_objective}
    \hat{\theta} = \arg\min_\theta ||Y - \theta||_2^2 + \lambda\sum_{t=1}^{n-1} |\theta_{t+1} - \theta_t|,
\end{align}
where $\lambda$ is a tuning parameter that controls the degree of regularization. We optimize $\lambda$ from the data as part of our overall fitting procedure based on BIC minimization (Section \ref{sec:bic_optimzation}). 
The estimator $\hat{\theta}$ will necessarily be a piecewise constant vector, due to the penalty term $\lambda\sum_{t=1}^{n-1} |\theta_{t+1} - \theta_t|$ that induces sparsity on the differences of $\theta$ \citep{tibshirani2005sparsity}.

It seems we could simply identify changepoints with the fused lasso by the breaks in $\hat{\theta}$, i.e., elements of the set $ A(\hat{\theta}) = \{t : \hat{\theta}_t \neq \hat{\theta}_{t+1} \}$ for a given $\lambda$.
However, the set $A(\hat{\theta})$ is sub-optimal for changepoint detection because the fused lasso tends to overestimate the number of changepoints and split single large breaks into multiple, tightly grouped, and smaller breaks \citep{fryzlewicz2014wild, rojas2014change}. Moreover, while the fused lasso is optimally adaptive to piecewise constant signals \citep{guntuboyina2017adaptive}, it lacks consistent support recovery \citep{rojas2014change}, meaning changepoint detection with $A(\hat{\theta})$ is inconsistent without further refinement \citep{chan2014group, hyun2016exact}. 

We propose a changeset regionalization procedure to further sparsify and improve the fused lasso estimates as a solution for changepoint detection. Recognizing that the fused lasso tends to estimate a single changepoint with a sequence of closely grouped estimates around the true value \citep{rojas2014change}, the regionalization procedure aims to collapse nearby estimates together and create a sequence of ``changesets" that likely correspond to only one changepoint. This allows the estimated changepoints to be reevaluated using a more powerful CUSUM statistics for each changeset and thus corrects the overestimation and inaccuracy issues of the fused lasso.

We first construct the changesets from the fused lasso estimated changepoints. We define the linkage parameter $c$ as the maximal distance any two neighboring estimated changepoints, which were supposed to correspond to a common actual change point, can be apart.
That is, we assume the true changepoints are at least $c+1$ time steps apart. Based on this assumption, we agglomerate the elements of $A(\hat{\theta})$ into sets, that are at least $c+1$ time steps apart, by linking together estimated changepoints that are within $c$ time steps of each other. Like $\lambda$, we optimize $c$ from the data as part of our overall fitting procedure based on BIC minimization (Section \ref{sec:bic_optimzation}).

To illustrate the agglomeration process, suppose we observe $Y_1,...,Y_{200}$ and the fused lasso estimates changepoints at the locations
\begin{equation*}
	A(\hat{\theta}) = \{10, 51, 54, 58, 100, 103, 106, 108, 112, 123, 140, 141\}.
\end{equation*}
We use our fitting procedure to estimate the optimal linkage parameter as $c = 5$. Agglomerating $A(\hat{\theta})$ with linkage parameter $c = 5$ yields the unique set of sets
\begin{equation*}
	B(\hat{\theta}, c = 5) = \{ \{10\}, \{51, 54, 58\}, \{100, 103, 106, 108, 112\}, \{123\}, \{140, 141\} \}.
\end{equation*}
We call the agglomerated changepoints the changesets and denote their collection as $B(\hat{\theta}, c)$.
Let $M = |B(\hat\theta, c)|$ denote the number of estimated changesets. Each changeset, denoted $b_j \in B(\hat{\theta}, c)$ for $j \in 1,...,M$, is intended to represent at most one true changepoint, since clusters of fused lasso based changepoint estimates tend to correspond to a single changepoint and we assumed that each true changepoint is at least $c+1$ time steps apart. For instance, the set $\{51, 54, 58\}$ represents the true changepoint at $t = 55$, while $\{100, 103, 106, 108, 112\}$ represents the true changepoint at $t = 107$.

The changesets in $B(\hat\theta, c)$ sparsify the fused lasso estimate, but they do not precisely estimate changepoints. To convert the changesets into changepoint estimates, we further introduce regionalization. Using the boundaries of the changesets, we split the data into regions assumed to contain AMOC each. 
We define the $j^{th}$ region, $R_j$, as the interval:
\begin{equation} \label{eqn:regions}
    R_j = [\sup b_{j-1} + 1, \inf b_{j+1} - 1],
\end{equation}
where $b_{j} \in B(\hat\theta, c)$ is the $j^{th}$ changeset, $\sup b_{0} = 0$ and $\inf b_{M+1} = n+1$.
Following the above example, $B(\hat{\theta}, c = 5)$ would generate the following regions
\begin{equation*}
	R_1 = [1, 50], R_2 = [11, 99], R_3 = [59, 122], R_4 = [113, 139], R_5 = [124, 200].
\end{equation*}
Each interval $R_j$ is the largest possible interval that contains $b_{j}$ and no other changesets. We, therefore, assume that each $R_j$ contains AMOC since the changeset $b_j$ is intended to represent a single changepoint.
Note that, since each changeset is at least $c+1$ timesteps apart, each region has at least $2c + 1$ data points. 

We then apply the standard CUSUM test region-wise to identify changepoint within each region. That is, we apply the CUSUM test to $Y_{R_j} = \{Y_{\sup b_{j-1} + 1,},...,Y_{\inf b_{j+1} - 1} \}$ for all $j \in 1,...,M$. Unlike the fused lasso, the CUSUM is a uniformly most powerful test for single changepoint detection and it is a consistent estimator of the changepoint location.
Because each region is tested independently, resulting in $M$ tests, we adjust the $M$ p-values using the Benjamini-Hochberg procedure \citep{benjamini1995controlling} to control the CUSUM's False Discovery Rate (FDR) at a prespecified level $\alpha$. We retain all estimated changepoints with an adjusted p-value below the prespecified $\alpha$ level.

\section{Simulations} \label{sec:simulations}
We investigate MCI's empirical performance in detecting changes in the mean and covariance of a functional process. We consider the case when there are no changepoints, when there are only a few changepoints (``sparse'' setting), and when there are many changepoints (``dense'' setting). We also consider functional data with light and heavy tails and with symmetric and skewed distributions. 

\subsection{Simulation setup}\label{sec:setup}

We start by simulating symmetric functional data. We use a Gaussian process (GP) model and a $t$-process (TP) model to simulate symmetric light-tailed and heavy-tailed functional data respectively. Let $Z_t$ denote the symmetric process. We have
$$Z_t = \mu_t + \epsilon, $$
where $\mu_t \in L^2([0, 1])$ is the mean and $\epsilon$ follows a zero-mean GP or TP with Mat\'ern covariance function
\citep{stein2012interpolation},
\begin{equation} \label{eqn:matern}
    C(x, x') = \frac{\sigma^2 \sqrt{\pi} r^{2\nu}}{2^{\nu-1}\Gamma(\nu + 1/2)} \left(\frac{\|x - x'\|}{r} \right)^\nu K_{\nu} \left(\frac{\|x - x'\|}{r} \right),
\end{equation}
with variance parameter $\sigma^2$, range parameter $r$, smoothness parameter $\nu$, and $K_\nu(\cdot)$ as a modified Bessel function of the 2nd kind of order $\nu$. If $\epsilon$ follows a TP, then an additional degrees of freedom parameter $df$ is required; we set $df = 3$ in all simulations.  We set the smoothness parameter $\nu = 1$ to ensure mean square differentiability of the sample paths so that the TVN projections are meaningful.

To generate a skewed functional process $\{Y_t, t \in 1,...,n \}$ on the domain $[0, 1]$, we let  
$$ Y_t = \log(1 + e^{Z_t(s)}). $$
We call the transformed Gaussian process and transformed $t$-process a log-sum Gaussian process (LS-GP) and a log-sum $t$-process (LS-TP), respectively.
All results presented in this section are based on LS-GP and LS-TP simulated data because many real datasets, particularly the atmospheric emitted radiance interferometer data we analyze here, are skewed. Also, the asymmetric data represent a more challenging situation for changepoint detection. Results under a GP or TP show similar patterns as the LS-GP and LS-TP simulations, but unsurprisingly all changepoints detectors improved under symmetric distributions with light tails. We defer the GP and TP results to Section B in the supplement. 

\begin{table}[]
\centering
\begin{tabular}{|l|l|}
\hline
Parameter   & Possible Values \\ \hline
$\mu$    & $\psi_1$, $\psi_2$, $\psi_3$, $\psi_4$, $\psi_5$ \\ \hline
$\sigma^2$ & 0.50, 0.66, 0.83, 1.00, 1.16, 1.33, 1.50, 1.66, 1.83, 2.00 \\ \hline
$r$      & 0.1, 0.2, 0.3, 0.4, 0.5, 0.6, 0.7, 0.8, 0.9, 1.0   \\ \hline
$\nu$      & 1.0   \\ \hline
\end{tabular}
\caption{Possible values for the LS-GP (LS-TP) parameters. $\mu$ indicates the mean function, while $\sigma$, r, and $\nu$ are scalar valued parameters in the Mat\'ern covariance function. The $\psi$ functions are taken from \cite{chiou2019identifying} and are defined in Equation \ref{eqn:psi_functions}.}
\label{tab:param_values}
\end{table}

Table \ref{tab:param_values} provides a list of candidate values for the mean $\mu$, the variance $\sigma^2$, and the range $r$ to take in our simulation. Specific parameter setting are described in Section \ref{sec:assess_results}.
The parameter $\nu$ is not varied because it acts on the process in nearly the same way as $r$.
We specify possible mean functions $\mu$, for the LS-GP or LS-TP, using the $\psi$ functions from the simulations in \cite{chiou2019identifying}:

\vspace{-1em}
\bgroup
\singlespacing
\begin{equation} \label{eqn:psi_functions}
\begin{aligned}
    \psi_1(t) &= 5t^2 - \exp(1 - 20t), \\
    \psi_2(t) &= 0.5 - 100(t - 0.1)(t - 0.3)(t - 0.5)(t - 0.9), \\
    \psi_3(t) &= \psi_2(t) + 0.8\sin(1 + 10 \pi t), \\
    \psi_4(t) &= 1 + 3t^2 - 5t^3 + 0.6\sin(1 + 10 \pi t), \\
    \psi_5(t) &= 1 + 3t^2 - 5t^3.
\end{aligned}
\end{equation}
\egroup

\noindent Differences between $\psi$ functions, as measured by the $L^2$ norm, yield different scales of change between the mean functions. The change from $\psi_1$ to $\psi_2$ is the largest, $\psi_3$ to $\psi_4$ is moderate, and $\psi_2$ to $\psi_3$ and $\psi_4$ to $\psi_5$ are both small. 
Additionally, these $\psi$ functions represent changes in both magnitude and the shape of the functional process.


To generate $M$ randomly spaced changepoints in either $\mu$, $\sigma$, or $r$ of an \textit{i.i.d.} LS-GP (or LS-TP) sequence, we first randomly sample $M+1$ parameter values and $M+1$ segment lengths (i.e., segment sample sizes). We generically denote the sequence of parameter values as $\theta_1,...,\theta_{M+1}$ and the sequence of segment lengths as $n_1,...,n_{M+1}$. Only one parameter is varied at a time, while the others are fixed at pre-specified values.
To generate the functional time series, we sample $n_1$ LS-GP's (or LS-TPs) with parameter $\theta = \theta_1$, then $n_2$ LS-GP's (or LS-TPs) with parameter $\theta = \theta_2$, and so on for the $M+1$ segments. Each segment is concatenated together so that the boundaries between segments represent changepoints in the functional time series. Parameter values are sampled from Table \ref{tab:param_values}, and sampling is done so that consecutive values are not the same. Segment lengths are samples from either a $\text{Unif}[500, 1000]$ or $\text{Unif}[5000, 10000]$ for the dense or sparse setting, respectively.

\subsection{Assessment criterion}\label{sec:assess}
We ran 500 simulations for each parameter setting and computed the Annotation error \citep{truong2019selective} and the Energy distance \citep{szekely2003statistics}, also called Energy error, between the true changepoints and the estimated changepoints. The Annotation error, widely used for assessing changepoints detection, measures the difference in the number of detected changepoints and the number of true changepoints. The energy error measures the distance between the set of estimated changepoints and the set of true changepoints. Let $X = \{X_1,...,X_n\}$ and $Y = \{Y_1,...,Y_m\}$ be two sets. The Annotation distance between $X$ and $Y$ is calculated as
$$ d_A(X, Y) = |n - m|, $$
and the Energy distance between two sets is calculated as
$$ d_E(X, Y) = \frac{2}{nm}\sum_{i = 1}^n\sum_{j = 1}^m |X_i - Y_j| - \frac{1}{n^2}\sum_{i = 1}^n\sum_{j = 1}^n |X_i - X_j| - \frac{1}{m^2}\sum_{i = 1}^m\sum_{j = 1}^m |Y_i - Y_j|.$$
A low Annotation distance (error) means that the algorithm consistently estimates the number of changepoints correctly, while a low Energy distance (error) means that the estimated and true changepoints are very similar.
In all simulation settings, we compare the different detectors on their ability to achieve low Annotation and low Energy errors.

Because the DSBE method proposed by \cite{chiou2019identifying} is the only multiple changepoint detection method that is computationally comparable to MCI, we compare our method to DSBE. Additionally, we include comparisons against three univariate, linear time changepoint detectors applied to the two projections of the TVN and the first functional principal component: PELT algorithm \citep{killick2012optimal}, the r-FPOP algorithm \citep{fearnhead2019changepoint}, and the WBS procedure \citep{fryzlewicz2014wild}.  This will allow us to learn the advantages of our detection method on top of the projections.

For PELT, we used the \texttt{cpt.meanvar()} function in the \texttt{changepoint} package with default settings and method = ``PELT''. For r-FPOP, we used the \texttt{Rob\_seg.std()} function from the \texttt{robseg} package with $L1$ loss and tuning parameter $\lambda = 5\log(n)$. The authors recommended $\lambda$ to scale with the log of the sample size and the multiplier 5 was found to have the overall best results. For WBS, we used the \texttt{wbs()} function, from the \texttt{wbs} package, with default settings and changepoints found via ssic.penalty (Strengthened Schwarz Information Criterion) minimization. For DSBE, we used the author's provided code with the number of changepoint candidates $K = 50$, minimum segment length $b = 5$, and significance threshold $\alpha = 0.05/K$. DSBE is only applied to the mean change simulations because DSBE was designed to only detect mean changes. 

Although we have tried to select the parameters to favor the above-mentioned methods, some default settings of those functions may not lead to optimal results. Nevertheless, We think it is still a fair comparison with our MCI method because we did not tune the parameters to favor our results. Instead, we estimated the $\lambda$ and $c$ parameters each time using a quick grid search (Section \ref{sec:bic_optimzation}). We set $\alpha = 0.001$ to keep the false positive rate low and avoid many false changepoints, similar to what was done for DSBE, which controls $\alpha = 0.05/K$.
It is a common practice for changepoint detection to control the false positive rate at a low level. 
We set $\alpha = 0.001$, but users can choose other small $\alpha$ values though extremely small $\alpha$ may lead to missing changepoints.  We conducted a small simulation study (Section B.2 in the supplement) to test the sensitivity of MCI to $\alpha$. We found that MCI's power is generally insensitive to $\alpha$ as long as $\alpha$ is small, say smaller than 0.001.

Note that we are not evaluating the three methods, PELT, FPOP, and rFPOP in the general settings for univariate time series. The results below only show their performance for projections of skewed and heavy tailed functional data, on which we focus.  Conclusions from the following simulations should not be extended to the general univariate setting.

\subsection{Assessment results} \label{sec:assess_results}
\subsubsection{No changepoints} \label{sec:none_sims}
\begin{figure}
  \centering
  \includegraphics[width=0.7\textwidth]{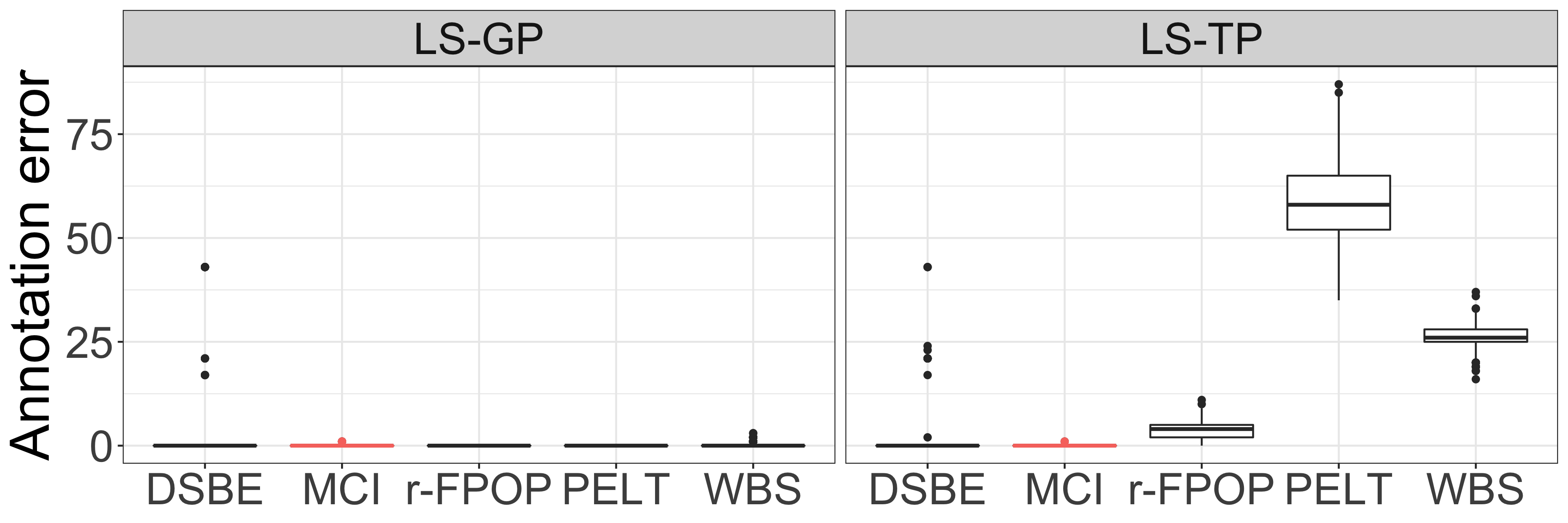}
  \caption{Annotation error when no changepoints are present in the data.
  MCI and DBSE nearly always detect 0 changepoints in both the light tail (LS-GP) and the heavy tail (LS-TP) setting. PELT and WBS detect an average of 55 and 25 changepoints, respectively, when no changepoints exist and the functional process is heavy tailed. FPOP performs better than PELT and WBS, but still detects around six false changepoints on average.}
  \label{fig:none_annotation}
\end{figure}

We first consider the situation where there are no changepoints, i.e. $M = 0$. Functional observations are generated from either a LS-GP or LS-TP with constant mean function $\mu = 0$ and Mat\'ern covariance with $\sigma = 1$, $r = 0.2$, and $\nu = 1$. Figure \ref{fig:none_annotation} shows distribution of Annotation errors under the LS-GP and LS-TP for each method.

All methods have an almost uniformly zero Annotation error under the LS-GP, meaning that all methods have an essentially zero false positive rate. However, under the LS-TP, only MCI and DSBE maintain their almost uniformly near-zero error rates, while the Annotation error for FPOP, PELT, and WBS all increases dramatically. PELT and WBS have particularly high Annotation error rates, possibly due to their sensitivity to larger random fluctuations caused by the heavy tailed generating process. FPOP seems more robust than PELT and WBS, likely because FPOP uses the robust $L1$ loss function, but it is still less robust than either DSBE or MCI.

\subsubsection{Sparse changepoints} \label{sec:sparse_sims}
We next consider the situation when the changepoints are relatively far apart, i.e., the changepoints are sparse. We simulate data as described in Section \ref{sec:setup} with $M = 5$ changepoints and segment lengths sampled from $\text{Unif}[5000, 10000]$. To study changepoints in $\mu$, we fix the covariance parameters to $\sigma = 1$, $\nu = 1$, and $r = 0.2$. For changepoints in variance, we fix $\mu = 0$, $\nu = 1$, and $r = 0.2$, and for changepoints in range we fix $\mu = 0$, $\nu = 1$, and $\sigma = 1$. 

\begin{figure}
  \centering
  \includegraphics[width=0.8\textwidth]{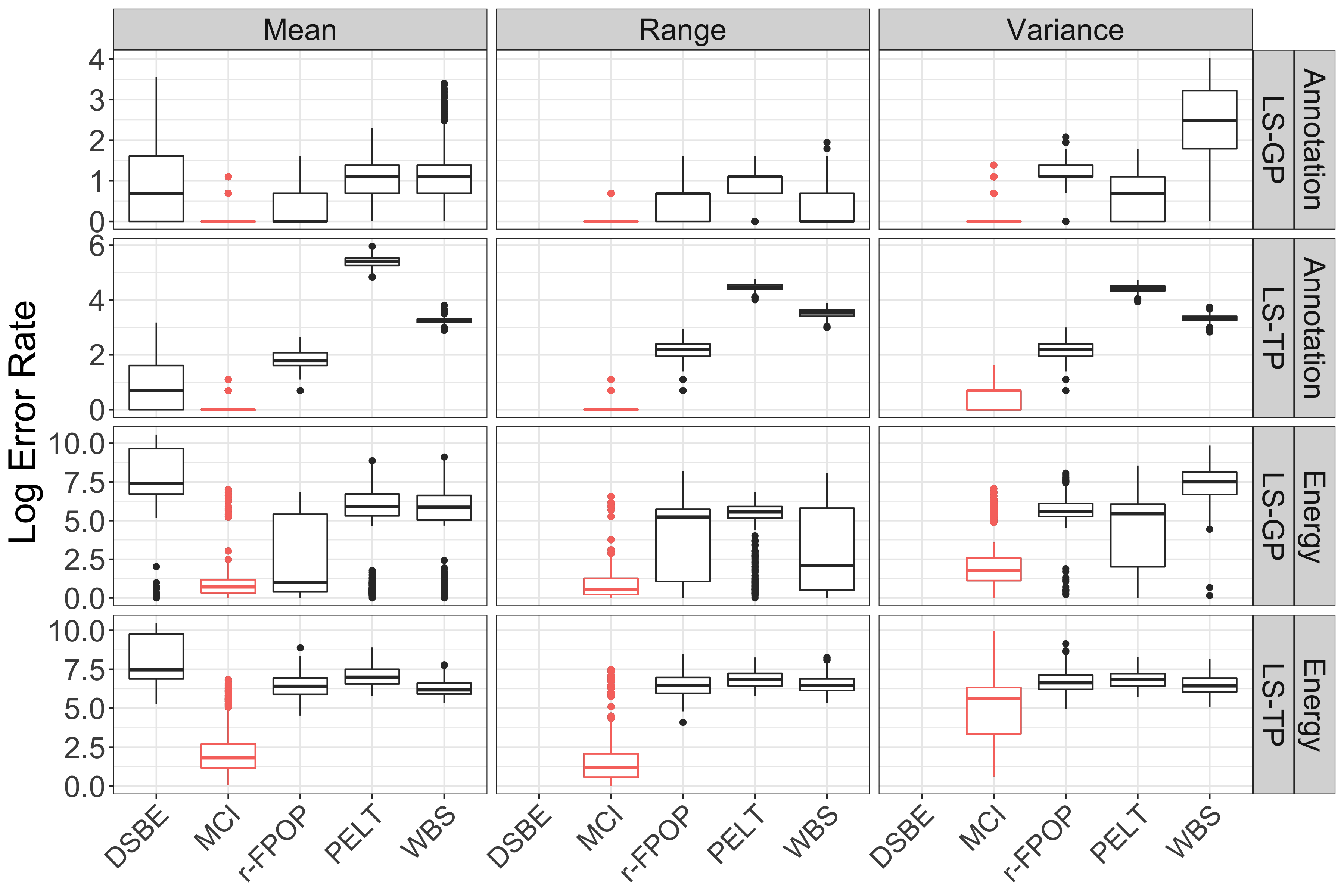}
 \caption{Annotation and Energy error rates for sparse changes in the mean function, range, and variance. Both error types are plotted on the $\log(1 + \text{error})$ scale. Under both the log-sum Gaussian process (LS-GP) and the log-sum $t$-process (LS-TP), the MCI method has far lower error rates than the other methods for detecting all changepoint types.}
  \label{fig:sparse_asym}
\end{figure}
 
Figure \ref{fig:sparse_asym} summarizes each method's Annotation error and Energy error in detecting changes in the mean, range, and variance. Our MCI method maintains the overall lowest Annotation error and the overall lowest Energy error. Together, this shows that MCI improves changepoint estimation in terms of both the number and location, whether the error process is light or heavy tailed. The univariate methods applied to our projections have reasonably low Annotation error rates on the LS-GP. However, they show high Annotation error rates on the LS-TP due to overestimation, while their Energy error rates are high on both. 
DSBE sees the most negligible deterioration from LS-GP to LS-TP compared to the alternative methods, although its Energy error rates were already high under LS-GP.

\subsubsection{Dense changepoints} \label{sec:dense_sims}

Finally, we consider the situation when the changepoints are relatively close to each other, i.e. when changepoints are dense. To simulate data with dense changepoints, we set $M = 50$ and sample segment lengths from $\text{Unif}[500, 1000]$. This design results in changepoints that are, on average, ten times as dense as the sparse changepoints in Section \ref{sec:sparse_sims}. The parameter setting for studying changes in the mean, variance, and range are the same as in Section \ref{sec:sparse_sims}. 

\begin{figure}
  \centering
  \includegraphics[width=0.8\textwidth]{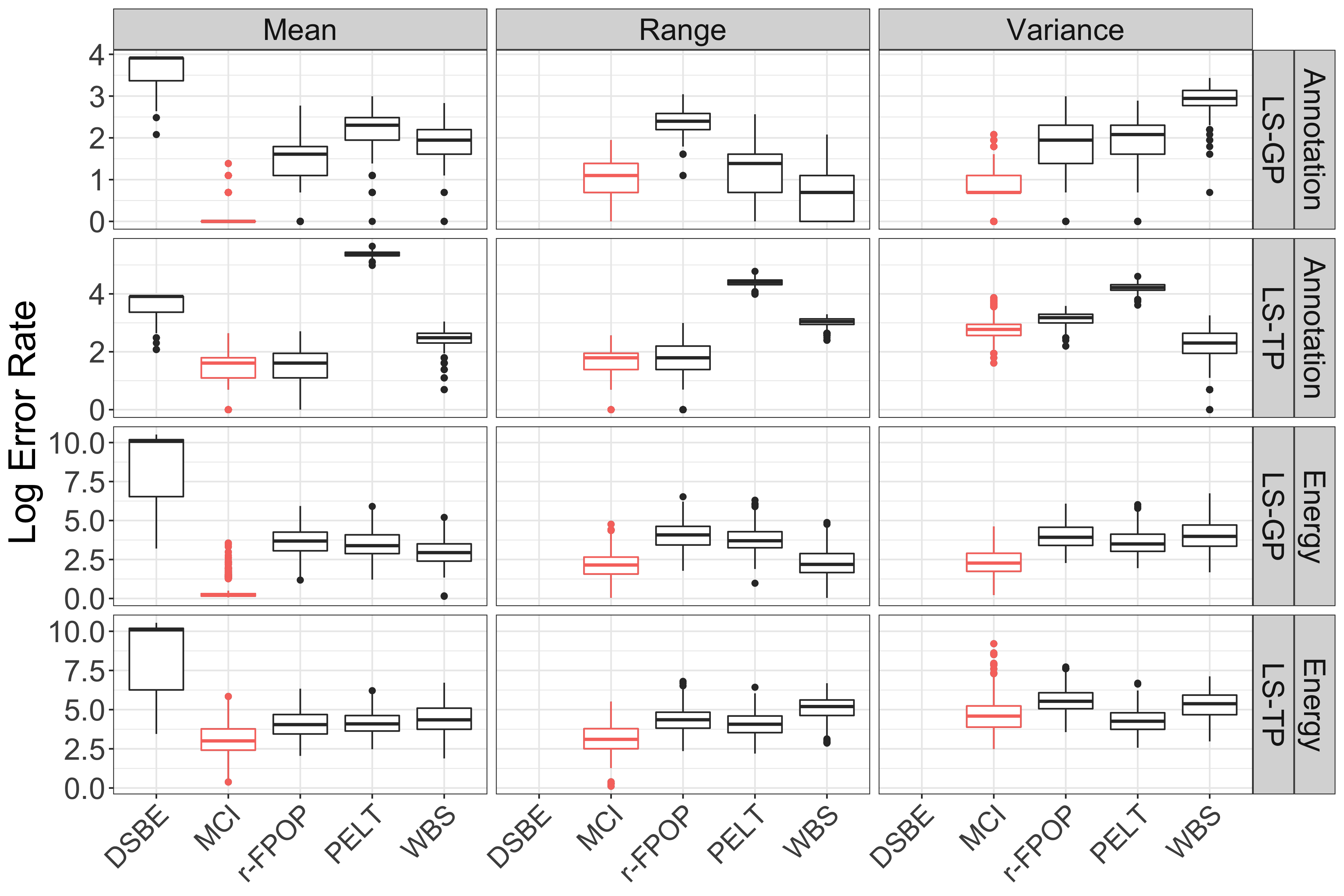}
  \caption{Annotation and Energy error rates for dense changepoints in the mean, range and variance parameters. Error values are plotted on the $\log(1 + \text{error})$ scale. MCI performs the best for detecting mean and variance changes in the LS-GP and for detecting range changes in the LS-TP. No method dominates for detecting variance changes in the LS-TP. }
  \label{fig:dense_asym}
\end{figure}

We summarize each method's ability to detect dense mean, range, and variance changes in Figure \ref{fig:dense_asym}. MCI again attains the lowest overall Annotation and Energy error rates. However, we find that the gap between MCI and other methods is smaller than when the changepoints were sparse. On the one hand, this is because MCI's error rates increase due to the CUSUM having fewer data per segment to estimate the location of changepoints. On the other hand, the competing methods see a decrease in their error rates for two reasons. One is that PELT and FPOP tend to estimate many changepoints, so when the actual number of changepoints is also high, their Annotation and Energy errors will naturally drop. The other reason is that WBS uses randomly sized intervals and a binary segmentation algorithm to find changepoints, both of which make the method less sensitive to changepoints density.

In summary, Figures \ref{fig:sparse_asym} and \ref{fig:dense_asym} show that MCI is generally more accurate and more skillful for a broader type of changepoints than the other approaches. This is because MCI effectively corrects the overestimation of fused lasso solution through changeset regionalization and application of CUMSUM. 
Under the light tailed LS-GP data, MCI has the lowest Annotation and Energy error rates across all changepoint types and densities. However, the three other algorithms' performance varies greatly depending on the changepoints' type and density.
With the heavy tailed LS-TP data, the error rate of MCI increases across all changepoint detection compared to LS-GP data. Nevertheless, the other methods show more deteriorated skill for heavy-tailed data.

\subsection{Linear computation}
Our method's computational complexity grows linearly with sample size because each sub-step of the method grows linearly in time, and the number of sub-steps does not grow with sample size.
For the FPC decomposition, we use the FACE algorithm \citep{xiao2016fast}, which is linear in sample size and function length. Computing the TVN of each function requires only a single pass over the data, so the TVN projection is linear as well. For fused lasso estimation, we use the Condat algorithm \citep{condat2013direct}, which is linear in time. Refining the segmentation over the data with changeset regionalization can be done with a single iteration. Finally, the CUSUM is computable in linear time, and the overlapping regions only cause CUSUM to be run (nearly) twice over the data set. The optimization procedure for 
$\lambda$ and $c$ uses a grid search over a fixed sequence of $\lambda$ and $c$ values, ensuring that the optimization step is also linear in sample size.

\begin{figure}
   \centering
  \includegraphics[width=0.7\textwidth]{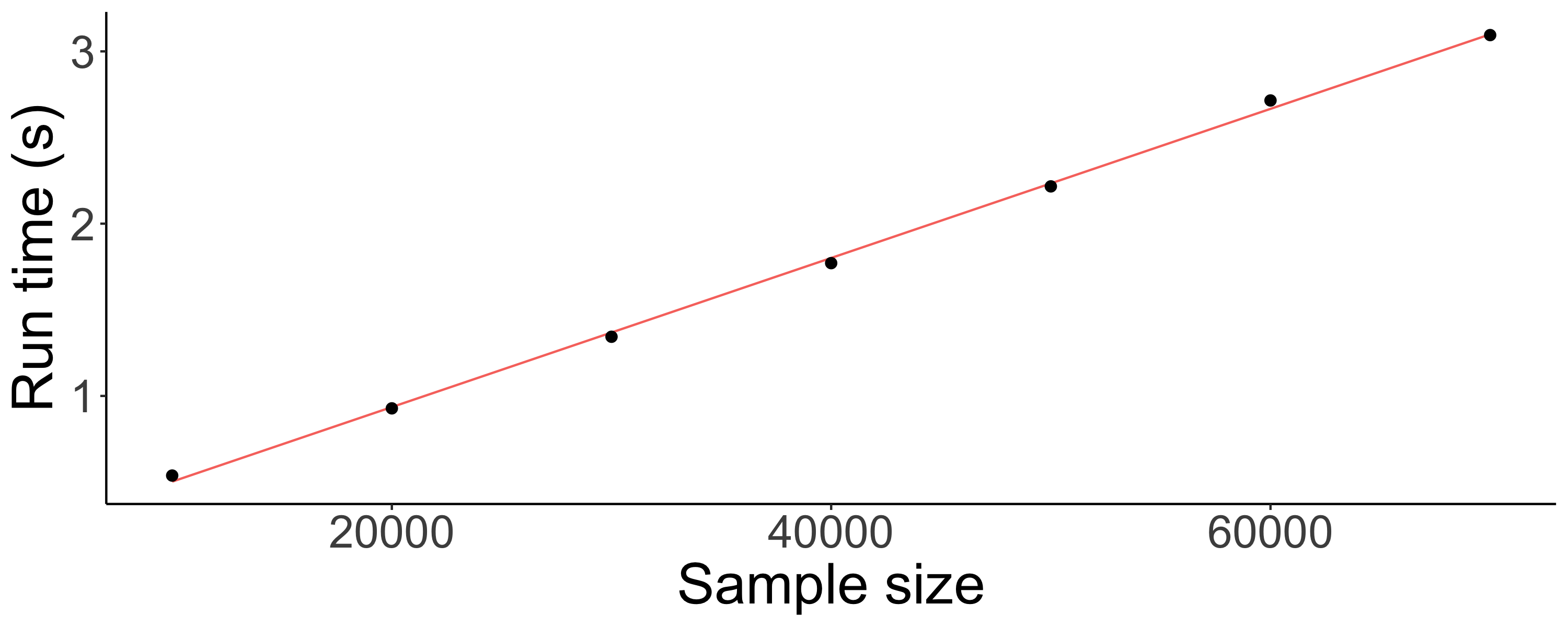}
  \caption{Median run time of MCI (black points) under seven sample size settings (10000-70000 observations). Trend line (red) shows the claimed linear relationship between sample size and computational time. Simulations were performed on single core of a 2017 MacBook Pro with a 2.33GHz i5 processor.}
  \label{fig:run_time}
\end{figure}

To demonstrate MCI's linear computation time, we conduct a simulation study to empirically assess the run time growth of MCI over an increasing sequence of sample sizes. We take the simulation with mean zero GP and covariance parameters $\sigma = 1$, $r = 0.2$, and $\nu = 1$ as example. We consider sample sizes from $n = 1000$ to $n = 7000$ in $1000$ unit increments, and run simulation 1000 times for each sample size.  We show the run time of MCI under each of the seven sample sizes in Figure \ref{fig:run_time}. The run time of MCI exhibits a linear trend with the sample size.

\section{Application to Profiles of Water Vapor} \label{sec:aeri}

The water vapor mixing ratio is the water vapor density over the dry air density in a given atmospheric unit. It is an important variable in Meteorology for distinguishing individual air masses, monitoring the effects of soil evapotranspiration and large water body evaporation \citep{north2014encyclopedia}, and for the early detection of heavy precipitation events \citep{sakai2019automated}. Changepoint detection helps identify sudden changes in an air parcel's water vapor content due to precipitation events and air parcels mixing. Retrospectively identifying sudden changes in the water vapor profiles' structure is often necessary before constructing statistical models for identifying precipitation events.

We apply our functional changepoint detector to the water vapor mixing ratio profiles collected from the Atmospheric Emitted Radiance Interferometer (AERI) instrument at the Lamont, Oklahoma Facility. The AERI instruments are maintained by the U.S. Department of Energy's (DOE) Atmospheric Radiation Measurement (ARM) Program to collect high-resolution atmospheric profile data \citep{stokes1994atmospheric}. The raw data are openly available in ``aeri01prof3feltz'' at \texttt{http://dx.doi.org/10.5439/1027271}. In this dataset, each profile consists of 58 measurements of the water vapor mixing ratio along a single atmospheric column from 0 to 44,000 meters above ground level in Lamont, Oklahoma. Complete profiles were collected every 8 minutes, thus providing near-continuous monitoring of atmospheric conditions. We removed the top 18 altitude points representing the 11,000 to 44,000-meter range due to the extremely high rate of measurement errors in this range. Therefore, we only consider the 40 measurements from 0 to 10,000 meters.



\begin{figure}
  \centering
  \includegraphics[width=0.9\textwidth]{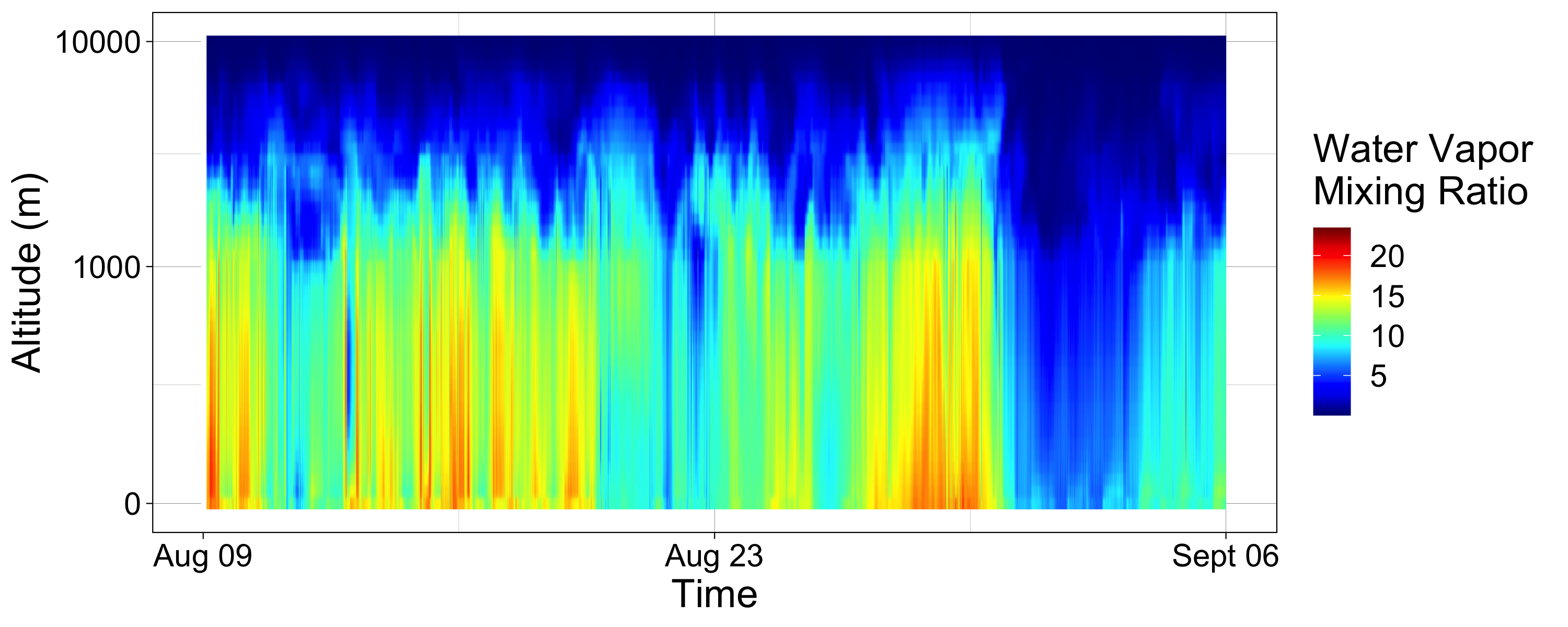}
  \caption{AERI water vapor mixing ratio profiles from August 9th to September 6th 2008. The $x$-axis is time, the $y$-axis is altitude, and the color values indicate the density of water vapor at each altitude and time. Each vertical line is an entire function.}
   \label{fig:aeri_examples}
\end{figure}

For our analysis, we consider the entire time series of water vapor profiles from January 4th, 2007 to March 10th, 2014. This period corresponds to 234,062 profiles, each sampled at the same 40 altitudes. 
To illustrate the data, we plotted profiles of water vapor mixing ratios from August 9th through September 6th in 2008 in Figure \ref{fig:aeri_examples}. 
Each vertical line represents an individual profile, with colors indicating the value of the profile at each altitude. Abrupt increases and decreases in water vapor along time are visible, indicating rapid changes from high density to low density and vice versa. The changes could be caused by sudden precipitation events or air mass mixing. We examined the marginal distribution of this data and found non-Gaussian behavior, including heavy tails and kurtosis.

\subsection{Identification of changepoints}

\begin{figure}
  \centering
  \includegraphics[width=0.9\textwidth]{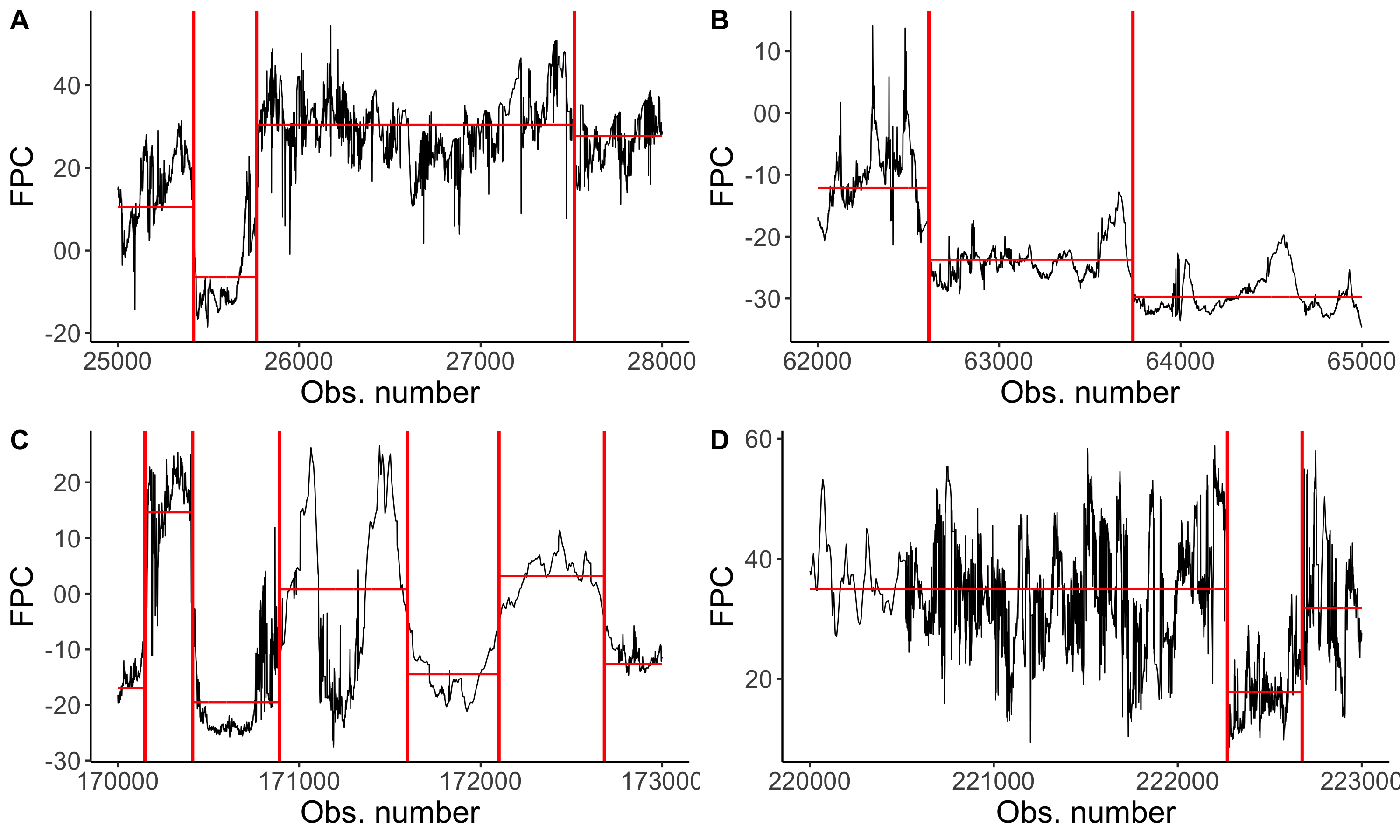}
  \caption{Four examples of the changepoints detected by MCI in the $P_{E1}$ projection of the AERI profiles. The black curve is the projected values, the red vertical lines mark the estimated changepoints, and the red horizontal lines indicate the segment means between changepoints}
  \label{fig:fpc_cp}
\end{figure}

\begin{figure}
  \centering
  \includegraphics[width=0.9\textwidth]{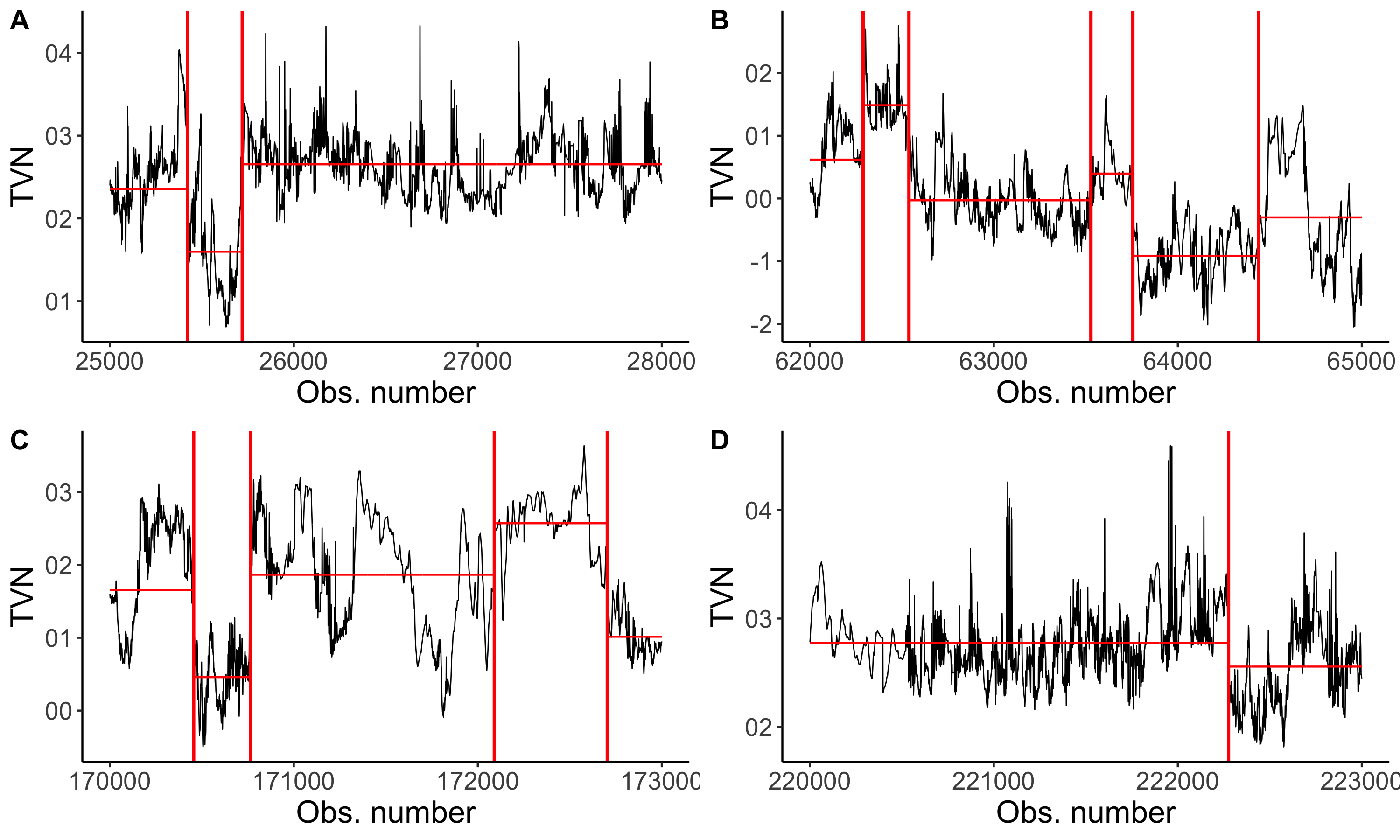}
  \caption{Four examples of the changepoints detected by MCI in the $P_{T}$ projection of the AERI profiles. The black curve is the projected values, the red vertical lines mark the estimated changepoints, and the red horizontal lines indicate the segment means between changepoints.}
  \label{fig:tv_cp}
\end{figure}

We applied our MCI method to the water vapor mixing ratio profiles and found 210 changepoints. Figure \ref{fig:fpc_cp} shows four examples of the changepoints identified in the first functional principal component projection ($P_{E1}$), and Figure \ref{fig:tv_cp} shows four examples of changepoints identified in the TVN projection ($P_{T}$). 

Several commonalities between the two plots are apparent. The first is that panels A and D in both figures are highly similar in appearance, and the changepoints identified in these regions are similar between the two projections. This happens because the $P_{E1}$ and $P_{T}$ are not necessarily orthogonal to each other, and changepoints may manifest in both low and high frequency spectrums. Another common feature is that MCI is robust to independence violations and heavy tails in $P_{E1}$ and $P_{T}$. This can be seen in the right half of panel C in both Figures \ref{fig:fpc_cp} and \ref{fig:tv_cp}, where both time series exhibit an autoregressive structure, yet MCI does not seem to yield overly dense changepoints that might be caused by the correlation.
Heavy tailed behavior can be observed in Panels A, B, and D, where large spikes in the time series are observed. MCI does not detect these anomalies as changepoints in either projection.

Figures \ref{fig:fpc_cp} and \ref{fig:tv_cp} also show that there are many differences between the two projections, meaning that $P_{E1}$ and $P_{T}$ measure very different aspects of the data. For example, panels B and C are almost entirely different in the two plots, including the locations of the detected changepoints. MCI was, therefore, able to pick up on a broader range of changepoints than $P_{E1}$ would allow because of the $P_{T}$ projection.

\section{Discussion}

We propose the Multiple Changepoint Isolation (MCI) method for detecting multiple changepoints of a functional time series. The changes can be either in the mean or in the dependence structure of the functional data. We first introduce a minimal system of projections to represent the variability ``between'' and ``within'' each function.
Motivated by the non-normal behavior of the TVN projection, we then introduce an augmented fused lasso based strategy to robustly segment the time series into regions likely containing at most one changepoint.
Finally, CUSUM is applied region-wise to detect and identify each region's potential changepoint.
Our extensive simulations show that the MCI method is accurate, computationally efficient, and robust to the underlying data distribution. Finally, we demonstrate MCI on water vapor mixing ratios over time.

Our two projections, the total variation norm and the first FPC, efficiently represent the major variability ``within'' and ``between'' each function, respectively. Our method is, therefore, able to detect a broad range of changepoints stemming from changes in the mean and covariance structure of the data.
In contrast, entirely functional metric based approaches are, in general, only powerful against changes in the mean \citep{aue2018detecting}. Our minimal data reduction, compared with using many FPCs, also has computational and theoretical benefits. The computational burden is lessened by only needing to compute a single FPC, and we conduct far fewer tests than if we had used numerous FPCs. Furthermore, conducting fewer tests helps our detector maintain higher power since more testing means harsher multiple testing corrections.

Our changepoint detection differs from the existing fused lasso based approaches in two significant ways. First, we used the fused lasso only as a segmentation procedure to identify regions likely to contain only at most one changepoint. Second, changepoint estimation is conducted via CUSUM testing and is not based on the specific jumps in the fused lasso estimate. This strategy is more efficient and powerful than directly using the fused lasso for changepoint detection since the fused lasso is shown only to be $\epsilon$-consistent \citep{rojas2014change} and CUSUM is the UMP test.

In future work, we would like to consider the theoretical underpinnings of the MCI method more rigorously. For instance, it remains to be shown whether the MCI is a consistent estimator of the changepoints or whether the MCI is asymptotically powerful, although our simulations seem to imply both properties. We may further study the MCI's robustness and compare its current form with alternatives using robust changepoint statistics rather than CUSUM. The procedure to optimize the tuning parameters $\lambda$ and $c$ (Section \ref{sec:bic_optimzation}) is also heuristic and not guaranteed to find globally optimal parameters. A better algorithm with optimality guarantees may be possible. However, we found through testing that the basin of optimal solutions is generally quite large, so more precise estimation algorithms may not be necessary. Finally, we may also consider extensions to higher dimensional functional processes, such as multivariate functional time, continuous surfaces, and spatial fields.

The supplement for this article is available online at the journal's website.


\baselineskip=20pt
\bibliography{refs}

\baselineskip=24pt
\appendix
\section{Appendix}
\subsection{CUSUM statistic} \label{sec:cusum_test}
The CUSUM statistic is a powerful and classical statistic for detecting a \textit{single} change in the mean of a univariate time series. \citep{page1954continuous, maceachern2007robust}. Let $Y = \{Y_t, t \in 1,\dots,n \}$ generically denote a univariate sequence. We can use the CUSUM statistic to test for a single change in the mean of the univariate sequence. That is, we can test the hypothesis that
\begin{align*}
    & H_0 : E(Y_1) = \dots = E(Y_n), \\
    & H_A : E(Y_1) = \dots E(Y_k) \neq E(Y_{k+1}) = \dots = E(Y_n), 
\end{align*}
for an unknown $1 < k < n$. A CUSUM process is defined as
\begin{equation*}
    T_n(\lfloor nr \rfloor) = n^{-1/2}\sum_{t = 1}^{\lfloor nr \rfloor} (Y_t - \Bar{y}_n), \quad r \in [0, 1],
\end{equation*}
where $\Bar{y}_n = n^{-1}\sum_{t = 1}^n Y_t$. The CUSUM statistic is defined as the supremum of the scaled CUSUM process
\begin{equation} \label{def:og_cusum}
   G_n = \sup_{r \in [0, 1]} |T_n(\lfloor nr \rfloor) / \hat{\sigma}_n|,
\end{equation}
where $\hat{\sigma}^2_n$ is a consistent estimator of the long run variance $\sigma^2 = \lim_{n \rightarrow \infty} n \text{var}(\Bar{X}_n)$. We approximate $\hat{\sigma}^2_n$ with $\text{Var}(Y - \hat{\theta})$, where $\hat{\theta}$ a non-parametric estimate of the mean of $Y$.
Under $H_0$, 
\begin{equation*}
     G_n \xrightarrow{D} \sup_{r\in[0,1]}|B(r) - rB(1)|,
\end{equation*}
where $B(\cdot)$ is a Brownian motion so $B(r) - rB(1)$ is a Brownian bridge on $[0, 1]$. Critical values for $\sup_{r\in[0,1]}|B(r) - rB(1)|$ can be computed using the well known Kolmogorov Distribution \citep{shao2010testing}:
\begin{equation} \label{def:cusum_pvals}
    P(G_n < t) = 1 - 2\sum_{j=1}^{\infty}(-1)^je^{-2j^2t^2}.
\end{equation}
If the test rejects $H_0$, then the estimated changepoint location is
\begin{equation} \label{def:cusum_cploc}
    \hat{k} = n \arg \sup_{r \in [0, 1]} |T_n(\lfloor nr \rfloor)|.
\end{equation}

The CUSUM estimator (\ref{def:cusum_cploc}) is highly powerful, in fact uniformly most powerful, for detecting a \textit{single} change in the mean of a univariate time series. However, in our formulation, we need to detect \textit{multiple} changes in the mean and covariance of functional data.

As mentioned in Section \ref{sec:intro}, existing strategies for multiple changepoint detection include dynamic segmentation and binary segmentation to augment CUSUM. However, these strategies may not be robust to heavy tailed or asymmetric data, such as our data (Figure \ref{fig:aeri_examples}). Another avenue for multiple changepoint detection, not relying on CUSUM, are the fused lasso \citep{tibshirani2011solution} based strategies \citep{rojas2014change, chan2014group, lin2016approximate, hyun2016exact}. These methods use the breaks or jumps in a fused lasso estimate to detect multiple changepoints. The fused lasso is more robust to heavy tails and anomalies, but its power is suboptimal compared with CUSUM, and it typically overestimates the number of changepoints \citep{chan2014group, fryzlewicz2014wild}.

In Section \ref{sec:methods} we will introduce a new strategy for multiple changepoint that combines the strengths of the CUSUM and fused lasso into a single procedure for detecting changes mean and covariance of functional time series.

\subsection{BIC optimization} \label{sec:bic_optimzation}

As mentioned in Section \ref{sec:methods}, we need optimize out the fused lasso penalty term $\lambda$ and the linkage parameter $c$. 
Following the methodology in \citep{chan2014group}, we consider Bayesian Information Criteria (BIC) minimization. Given a set of estimated changepoints $\tau_1,...,\tau_m$ we define $\theta^*$ as
\begin{equation*}
	\theta^*_t = \sum_{i = 0}^M \mathbbm{1}(\tau_i < t \leq \tau_{i+1}) \left[ \frac{1}{\tau_{i+1} - \tau_i} \sum_{t = \tau_i}^{\tau_{i+1}}Y_t \right],
\end{equation*}
where $\tau_0 = 1$, $\tau_{M+1} = n$, and $n$ is the sample size. We define $S_n(\tau_{i-1}, \tau_{i}) = \sum_{t = \tau_i}^{\tau_{i+1}} (Y_t - \theta^*_t)^2$ and $S_n(\tau_1,..., \tau_m) = \sum_{i = 1}^{m} S_n(\tau_{i-1}, \tau_{i})$. Finally we can define the BIC of the estimated changepoints as
\begin{equation*}
	BIC(\lambda, c) = S_n(\tau_1,...,\tau_m) + m \log(n),
\end{equation*}
and estimate $\lambda$ and $c$ by solving
\begin{equation} \label{eqn:bic_minimizaion}
	\hat{\lambda}, \hat{c} = \arg\min_{\lambda, c} BIC(\lambda, c).
\end{equation}
Minimizing Equation \ref{eqn:bic_minimizaion} is a proxy for finding $\theta^*$ with a minimal $L^0$ norm \citep{louizos2018learning}, or, equivalently, a minimal set of changepoints, $\tau_1,...,\tau_m$, sufficient to explain the variation in the projected data. 

The objective in Equation \ref{eqn:bic_minimizaion} is non-differentiable, so we use grid search to brute force estimating $\hat{\lambda}, \hat{c}$.
We restrict our search to multiples of  $\sqrt{n}$ for each parameter. That is, we allow each parameter to grow as a multiple of the square root of the functional time series' length, $\lambda = r \sqrt{n}$ and $c = k \sqrt{n}$. In our implementation we allow $r$ and $c$ to range from $0.2-5$ in increments of $0.1$.

The $\sqrt{n}$ scaling for $\lambda$  was used in \cite{lin2016approximate} for their fused lasso estimator to ensure the solution is well estimated while still filtering out a majority of the extraneous changepoints. \cite{rojas2014change} showed that $c$ depends on sample size $n$ as $c= kn$. However, directly using $n$ made the numerical estimation of $c$ challenging due to the small magnitudes of $k$. Instead, we made c scale with $\sqrt{n}$ and found this to work better in our simulations.




\end{document}